\documentclass[preprint,12pt]{elsarticle}

\usepackage{amssymb}
\usepackage{color}
\usepackage{tabularray}
\usepackage{booktabs}
\usepackage{makecell}
\usepackage{amsmath}
\usepackage{xurl}
\usepackage{graphicx}
\usepackage[bf]{subfigure}

\usepackage{microtype}
\usepackage{graphicx}
\usepackage{caption}
\usepackage{subcaption}

\usepackage{mathtools}
\usepackage{amsthm}
\usepackage{colortbl}

\usepackage{lineno}

\usepackage{etoolbox}
\makeatletter
\def\ps@pprintTitle{%
 \let\@oddhead\@empty
 \let\@evenhead\@empty
 \let\@oddfoot\@empty
 \let\@evenfoot\@empty
}
\makeatother
\begin{document}

\begin{frontmatter}



\title{Function and form of U.S. cities}


\author[inst1]{Sandro M. Reia}
\ead{smreia@gmail.com}
\author[inst1]{Taylor Anderson}
\author[inst1]{Henrique F. Arruda}
\author[inst1]{Kuldip S. Atwal}
\author[inst1]{Shiyang Ruan}
\author[inst2]{Hamdi Kavak}
\author[inst1]{Dieter Pfoser}

\affiliation[inst1]{organization={Geography and Geoinformation Science, College of Science, George Mason University},
            addressline={4400 University Dr.}, 
            city={Fairfax},
            postcode={22030}, 
            state={VA},
            country={U.S.}}

\affiliation[inst2]{organization={Center for Social Complexity, College of Science, George Mason University},
            addressline={4400 University Dr.}, 
            city={Fairfax},
            postcode={22030}, 
            state={VA},
            country={U.S.}}

\begin{abstract}
The relationship between urban form and function is a complex challenge that can be examined from multiple perspectives. 
In this study, we propose a method to characterize the urban function of U.S. metropolitan areas by analyzing trip patterns extracted from the 2017 National Household Travel Survey (NHTS). To characterize urban form, we employ measures that capture road network topology. We cluster cities based on both form and function and subsequently compare these clusters.
Our analysis of 52 U.S. metropolitan areas identifies 7 distinct clusters of cities that exhibit similar travel behavior, suggesting that diverse mobility patterns can be effectively grouped into a few universal classes.
The observed disparity between the urban-function clustering and the urban-form clustering suggests that travel behavior in the U.S. is not strongly influenced by the physical infrastructure of the city.
\end{abstract}







\end{frontmatter}


\section{Introduction}
\label{sec:sample1}

Cities are spatially complex and heterogeneous systems \cite{allen1997cities, bettencourt2015cities}.
The complex dynamics of city growth have led to non-uniform urban morphologies, often characterized by sparse populations and fractal-like geometries \cite{batty1989urban, batty1991cities, frankhauser2004comparing}.
In fact, it was recently shown that the population growth of cities is driven by asymmetric migratory shocks \cite{verbavatz2020growth, bettencourt2020demography}, so that the population growth of one city is often sustained by the population loss of others.
On a more granular level, migration also plays an important role in describing the intra-city spatial heterogeneity; population growth in core areas of cities is more significantly influenced by inter-city migration flows, while population growth in external areas is more heavily impacted by intra-city outflows from core areas \cite{reia2022spatial, reia2022modeling}.

While migration captures long-term mobility, trip chains capture urban mobility in a 24-hour time scale \cite{primerano2008defining}.
A trip chain is a sequence of trip segments beginning and ending at home \cite{primerano2008defining}.  Trip chains carried out by individuals within a city can be used as a proxy to describe its \textit{urban function}, a term referring to activities that take place within a city \cite{hu2021modeling}.
\textit{Urban form}, on the other hand, is related to the spatial structure of a city, capturing diverse aspects, such as landscape, economic structure,transportation, community design, and urban design \cite{clifton2008quantitative}.

There is a complex interplay between urban form and function. Urban function follows form, where the built environment shapes mobility and activity within a space  \cite{crooks2015crowdsourcing, dascher2019function}. At the same time, form follows function, meaning the activities within a space are thought to drive the emergence of form in urban environments \cite{batty1992form}. A well-planned city, one that balances form and function, can increase accessibility, reduce congestion, and promote sustainable living by integrating efficient public transportation systems. 
Conversely, poorly planned cities, such as sprawling suburbs, can lead to car dependency, increased pollution, and social isolation \cite{duany2000suburban}.
Despite the existing literature on the relationship between urban form and function \cite{burger2012form, crooks2015crowdsourcing, van2011mapping}, limited data means that it can be challenging to characterize the function of cities. As such, most research in this area has focused on specific case studies or a limited number of cities, constraining the generalizability of their findings. 

In this paper, we address this research gap by conducting a systematic analysis that explores the relationship between urban function and urban form in 52 metropolitan statistical areas (MSAs) in the U.S. While approaches to describing urban form of cities are well-established, we propose a framework that uses the 2017 National Household Travel Survey (NHTS) to characterize the urban function of the MSAs \cite{FHWA2022}. 
Note that we use the terms ``city'' and ``MSA'' interchangeably. 
First, based on travel behaviors captured by the NHTS data for each city, we cluster cities by their function, where cities within the same cluster have similar mobility patterns. 
This clustering suggests that complex human behaviors driving mobility in U.S. cities can be categorized into a few universal classes. 
Next, using Crucitti's network centrality measures \cite{crucitti2006centrality}, 
we cluster the cities by their urban form. 
Our findings indicate a lack of a clear correspondence between structural and functional clusters, suggesting that the function of these cities is less so shaped by the urban environment, and may instead be influenced more by cultural and population-specific needs \cite{maslow1987maslow}.



The paper is organized as follows: In the next section, we review the literature on the interplay between urban form and function. This is followed by Section~\ref{sec:methods}, where we detail the datasets utilized to characterize urban form and function, describe our mapping scheme, and explain the centrality measures employed. Our findings are then presented in Section~\ref{sec:results}. 
We conclude the paper by summarizing our key insights, discussing their potential generalizability and highlighting future research directions in Section~\ref{sec:concluisons}.

\section{Related Works}
\label{sec:related_works}

The study of urban form and function spans multiple disciplines, providing diverse insights into the dynamics of urban regions. Network science has emerged as a powerful tool for characterizing urban landscapes, with studies using network measures to delineate urban regions and quantify traffic flow along their streets~\cite{boccaletti2006complex,costa2007characterization}. In addition, analyses of human mobility often elucidated through trip chains and related data, provide a rich source of information for studying mobility patterns in urban environments \cite{alexander2015origin, wang2018applying,holguin2005observed, primerano2008defining, mcguckin2004trips, chauhan2021database}. There are two main ways of extracting these chains from a population: via a sequence of stay points captured with mobile phone data \cite{alexander2015origin, wang2018applying}, and via origin-destination trip information from travel surveys \cite{holguin2005observed, primerano2008defining, mcguckin2004trips, chauhan2021database}.

The extraction of chains from mobile phone data, despite offering high-resolution temporal and spatial information on individual movements, depends on the stay point inference methods, which are used to determine where the individuals are at any time of the day \cite{jiang2013review, wang2017trip}. 
Travel surveys, on the other hand, provide broader insights into travel patterns\cite{safi2017empirical}. 
Travel surveys are available for urban areas in many countries \cite{schafer2000regularities}; for example Australia \cite{stopher2011national}, France \cite{NTTS_2009}, Great Britain \cite{NTS_2023}, the Netherlands \cite{cbs_dutch_national_travel_survey}, Norway \cite{ssb_norwegian_travel_survey}, Switzerland \cite{bfs_swiss_travel_survey}, Austria \cite{statistik_at_austrian_travel_habits_survey},  Canada \cite{statcan_national_travel_survey} and the U.S. In the U.S., the National Household Travel Survey \cite{FHWA2022}, provides insights into the travel behavior the U.S. population. Respondents of the survey are asked to report their activities in a 24-hour time window, and it includes daily non-commercial travel by all modes \cite{FHWA2022}.


The analyses of travel patterns have the potential to provide insights into \textit{urban function}. 
It has been shown that human trajectories have a high degree of regularity, meaning that people tend to visit their preferred locations more often \cite{gonzalez2008understanding}.
This regularity in daily mobility patterns has also been captured in network analyses, which showed that more than $90\%$ of trip chains are well described by only $17$ unique motifs \cite{schneider2013unravelling}.
Short trip chains, such as home-work-home, home-education-home, and home-religious activity-home, are the most frequent chains \cite{ectors2019exploratory}.
Indeed, the statistical structure of trip chains and the prevalence of popular trip chains are well captured by Zipf's law \cite{ectors2019exploratory}, indicating that complex human behavior can be summarized by a simple power law structure.

\textit{Urban form} refers to different aspects of spatial organization and structure within cities and is characterized by different analytical frameworks \cite{vzivkovic2020urban, kropf2009aspects}.
For example, Crucitti \textit{et al.} \cite{crucitti2006centrality} characterizes urban form using a set of centrality measures obtained from a spatial network representing its road network infrastructure. 
Centrality is a key concept in complex network analysis, identifying the importance or influence of a node within a network. 
In urban studies, this concept helps us understand how cities function by examining their road networks. 
Urban street patterns can be analyzed as networks of roads (edges) connecting intersections (nodes), a representation that has been approached in various ways \cite{hillier1989social, jiang2004topological, porta2006network}. 
This framework distinguishes cities in terms of their structure and is less data and computationally intensive than counterparts, such as remote sensing-based analysis \cite{mesev1995morphology, herold2003spatial, wang2023eo+}, thus allowing comparison and classification of a large set of cities when data is scarce.

The principle of ``form follows function'' implies that the physical layout and structure of a city should be shaped by its intended purpose and activities \cite{lynch1964image, jacobsdeath}. 
For example, residential neighborhoods are typically designed with considerations for housing density, green spaces, and proximity to schools and amenities, reflecting their role in providing livable and accessible environments for families. 
Conversely, commercial districts are characterized by higher building densities, accessible transportation networks, and infrastructure that supports economic activities \cite{lynch1964image, jacobsdeath}.

At the same time, urban form enables and shapes activities within such areas. For example, the configuration streets, the placement of public transportation, and the distribution of amenities can determine travel patterns, social interactions, and economic activities. 
Polycentric cities, which have multiple centers of activity rather than a single center, can increase both economic productivity and environmental sustainability by reducing the need for long commutes and promoting diverse land use patterns \cite{burger2012form}. 
As cities grow, they tend to transition from a monocentric to a polycentric structure once congestion reaches an upper limit dictated by the city's population and road network infrastructure \cite{louf2013modeling}.

The form and function of cities can be mapped by surface maps that reveal the spatial distribution of impervious surfaces — areas covered by materials such as concrete, asphalt, and buildings that prevent water infiltration into the soil \cite{van2011mapping}. 
This approach captures the intricate patterns of built and non-built areas, providing critical insights into the urban landscape. 
However, the form and function of cities can also be assessed by how people define urban space through their activities \cite{crooks2015agent}. 
Crowdsourced data from social media, GPS devices, and other sources provide real-time insights into how people interact with urban spaces, revealing patterns of urban mobility and urban usage that traditional data sources might miss \cite{crooks2015agent}.

\section{Data and Methods}
\label{sec:methods}

\begin{figure}[h!]
\centering
\includegraphics[width=1\linewidth]{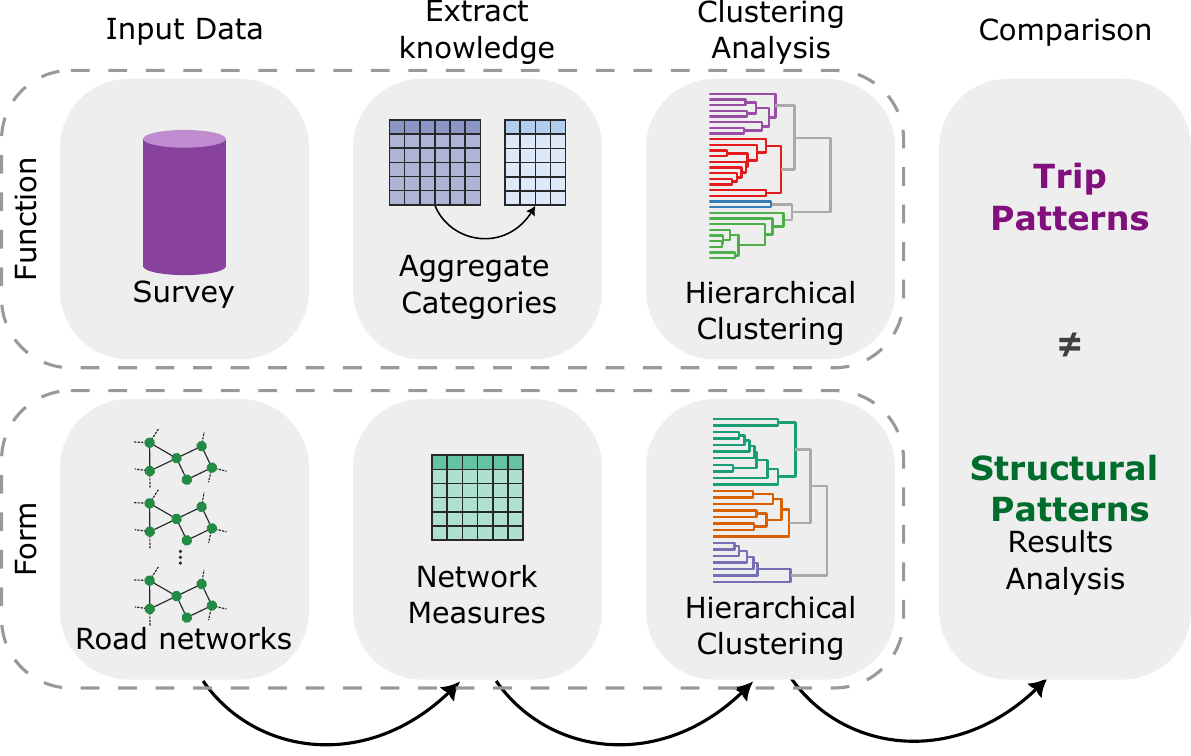}
\caption{\textbf{Schematic representation of our analysis.
}
First, we characterize the urban function of cities using information extracted from the NHTS data.
Second, we characterize urban form of cities by extracting network measures from spatial networks that capture road infrastructure. 
Finally, we compare the different clusters of cities based on urban form and function.
}
\label{fig:scheme}
\end{figure}

Here, we propose a framework for characterizing the urban function of U.S. cities based on mobility patterns captured by the NHTS.
This characterization allows us to cluster cities into $7$ functional classes, revealing cities with similar mobility patterns.   
We also characterize cities in terms of their urban form.
The urban form of a city is characterized using the framework proposed by Crucitti \textit{et al.} \cite{crucitti2006centrality}, where the structure of cities is captured by a set of four road network centrality measures.
Figure~\ref{fig:scheme} provides an overview of the analysis we conduct here.
In the following, we provide details on the datasets and methods we used to assess the urban form and function of the cities. 

\subsection{Urban function}

\subsubsection*{Input Data} 
The NHTS \cite{FHWA2022} provides open-access data on the travel behavior of the U.S. population. The data collected includes a record of all trips taken within a 24-hour period by all household members aged 5 or older \cite{FHWA2022}. The NHTS defines a ``trip'' as movement from one location (an origin) to another location (a destination) for a specific purpose on a respondent's travel day. There are $20$ different trip purpose categories captured by the survey including ``work'', ``attend school as a student'', ``buy goods'' etc. (Table \ref{table_survey_mapping}). Here, we distinguish a single ``trip'' from a trip chain, which is a sequence of trips visited in a 24-hour time period.

\subsubsection*{Knowledge extraction}

We follow the methodology outlined in \cite{balac2021synthetic} and first aggregate the 20 NHTS purpose-based categories to 11 aggregated activity-based categories (see Table \ref{table_survey_mapping}). This reduces the number of overall trip categories for analysis and provides more robust samples of less popular trips. For example, ``Home'' and ``Work from Home'' are aggregated into the new category ``Home'' due to the lack of travel involved, meaning there is no trip. In other cases, NHTS purpose-based categories that are similar are combined into one activity-based category to be used in our analysis. For example, we aggregate the original purpose-based categories ``Attend child care'' and ``Attend adult care'' into our new activity-based category ``Care.''

\begin{figure}[!ht]
\centering
{\includegraphics{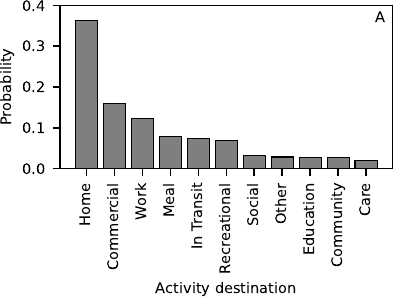}}
{\includegraphics{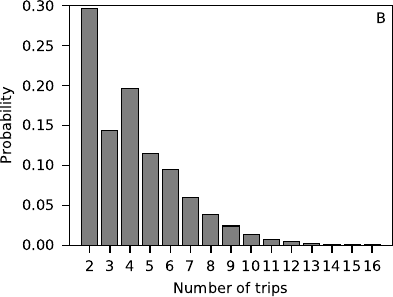}}
{\includegraphics{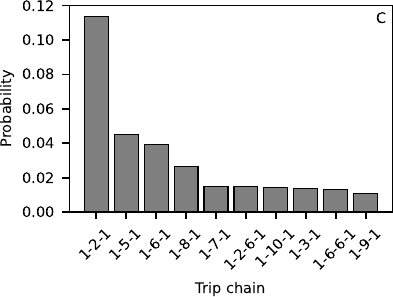}}
{\includegraphics{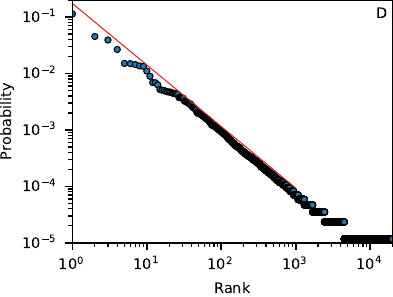}}
\caption{\textbf{Characterizing trip patterns based on the proposed set of activity categories}.
In panel A, the frequency of visits to ``home'' is more than twice that of visits to ``commercial'', which is the second most visited location.
In panel B, trip chains consisting of $2$ trips are the most common, accounting for almost $30\%$ of all trip patterns.
They are followed by chains of $4$ trips, corresponding to about $20\%$ of trip patterns. 
Panel C shows the probability of finding the top $10$ most common trip chains (each with a probability greater than $1\%$).  
These chains are denoted by activities labeled as 1 = Home, 2 = Work, 3 = Community, 4 = In Transit, 5 = Education, 6 = Commercial, 7 = Meal, 8 = Recreational, 9 = Care, 10 = Social, 11 = Other. 
Panel D illustrates the probability of finding an activity chain based on its frequency rank. The red line represents the function $Probability \propto 1 / rank^a$ with $a = 1.08$, suggesting that the probability distribution of trip chains follows Zipf's law.}
\label{fig:probability_activity_destination}
\end{figure}

Across all cities, ``home'' is the most popular destination, making up about $35\%$ of all trips (Figure \ref{fig:probability_activity_destination} A).
The high prevalence of trips to ``home'' suggests that the home is an activity hub, a central stay point in long trip chains.
Trips to commercial places (e.g. shopping) are the second most popular, making up about $17\%$ of all trips.
Not surprisingly, trips to work are the third most popular, where work is the destination in about $14\%$ of all trips. 
The frequency distribution of trips (Figure \ref{fig:probability_activity_destination} A) can be considered a proxy for the different needs of individuals in a city.
A higher frequency of trips to commercial places might indicate that the economy of that particular area is more service-centered. 
In contrast, high visits to ``care'' might indicate populations are dependent on others.

Recall that sequences of trips form a trip chain. Across all cities, short activity chains are the most predominant. About $30\%$ of the chains are composed of only $2$ trips (Figure \ref{fig:probability_activity_destination} B); For example, from home to some activity (trip 1) and then from that activity back to home (trip 2). The most common chain, accounting for nearly $12\%$ of all recorded trip chains, is the ``home-work-home'' sequence. This chain is about three times more common than the second most popular one, ``home-education-home''.
Surprisingly, chains with $4$ trips are also popular. These chains capture travel to a meal during lunch break (home-work-meal-work-home) or home-centered chains (home-work-home-commercial-home). The probability of finding a trip chain decreases as the length of the chain increases. As such, it is unlikely to find large trip chains (more than $16$ trips).
The disparity in the frequency of different trip chains is also illustrated by the rank plot  (Figure \ref{fig:probability_activity_destination} D), showing that the ranking of trip chains closely follows Zipf's law, confirming the asymmetric distribution of different trip chains, which indicates a preferential behavior towards some types of activities\cite{ectors2019exploratory}..
The adherence to Zipf's law is found in both the original travel survey data and our aggregated data, suggesting that we were able to maintain the structure of data even as we combined types of activities together.

For the purpose of our analysis, we extract the trip chains that start and end at home for each city and decompose each chain into its set of trips \cite{primerano2008defining}.
Next, for each city, we construct a $11\times11$ O-D matrix where each element of this matrix captures the frequency of trips from one activity to another. 

\begin{figure}[h]
\centering
    {\includegraphics[width=0.495\linewidth]{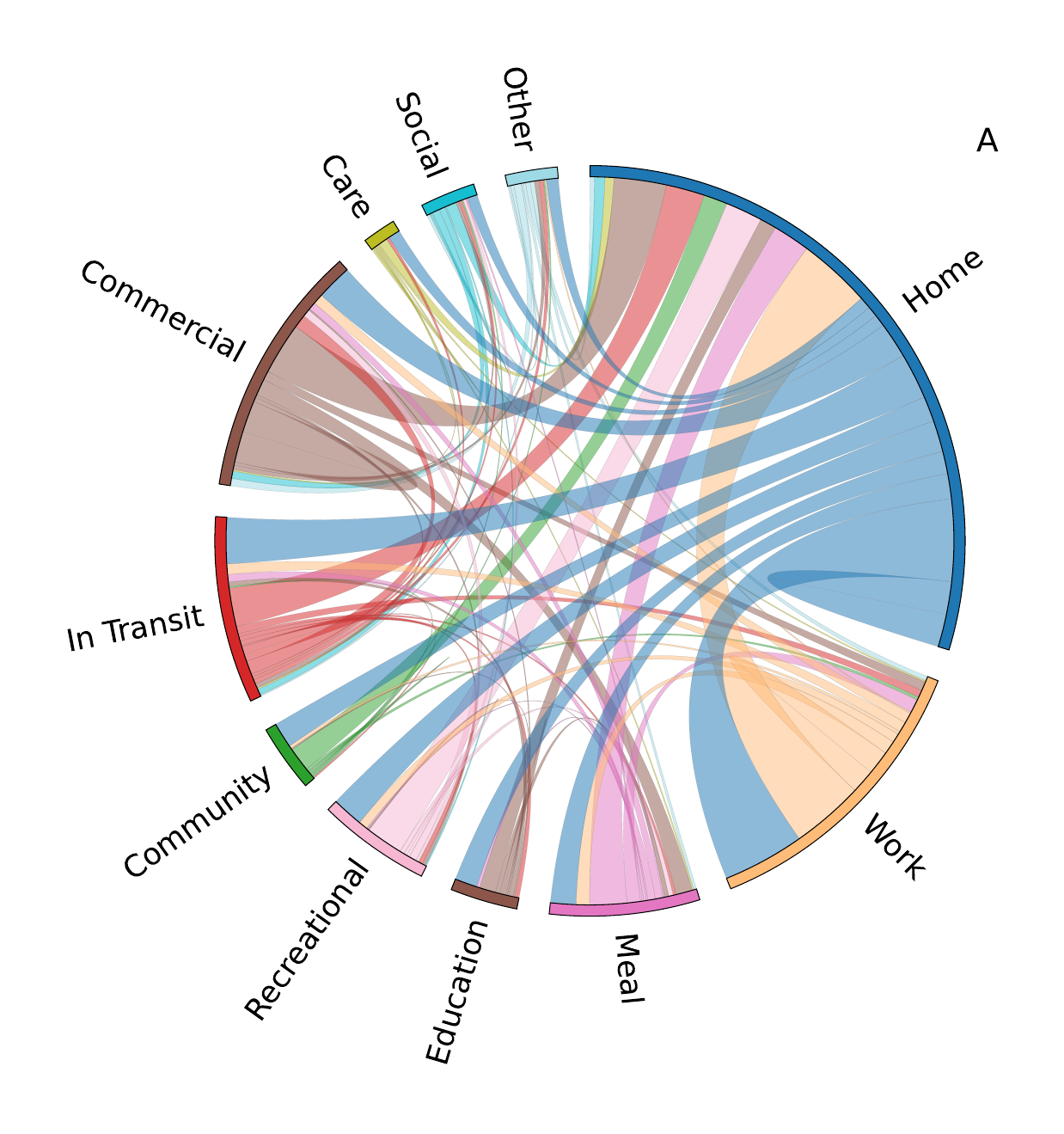}}
    {\includegraphics[width=0.495\linewidth]{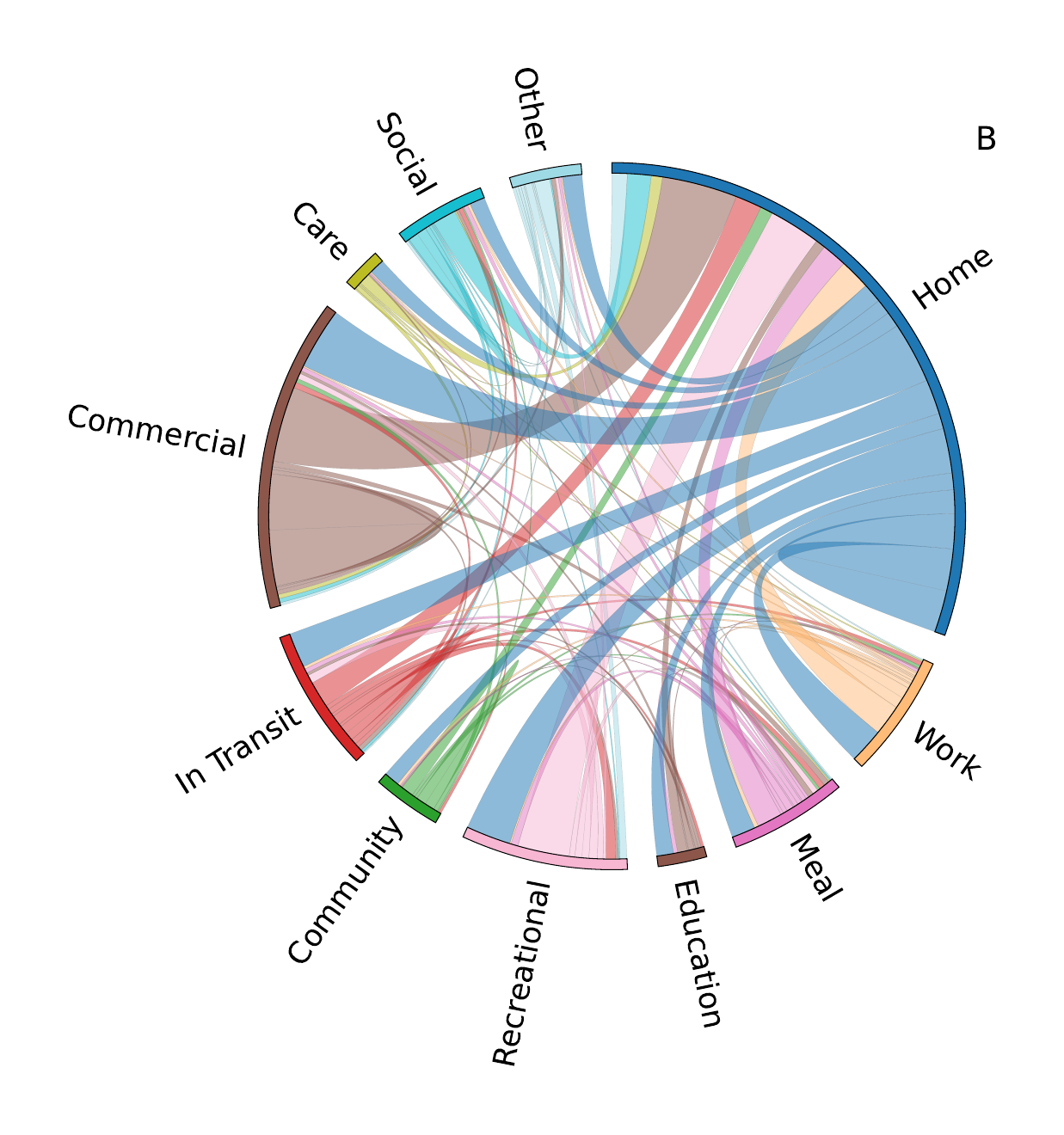}}
\caption{\textbf{Trip flow diagrams capture differences in travel patterns between cities}. 
The trip flow diagrams shown here illustrate the frequency of trips between different origin-destination activity-based category pairs. 
The starting points of the arrows represent the origins, while the endpoints represent the destinations. 
The width of the arrows is proportional to the frequency of trips for each specific OD pair. 
In panel A, the trip flow diagram for Grand Rapids-Wyoming, MI, shows a significant volume of trips originating or ending at Work. 
Conversely, panel B shows Hartford-West Hartford-East Hartford, CT, where there is a noticeable decrease in work-related trips but an increase in trips related to Commercial, Recreational, and Social activities.}
\label{fig:flow_diagram}
\end{figure}

The matrix for each city can be visualized as a trip flow diagram. For example, Figure \ref{fig:flow_diagram} compares the trip frequencies for two similarly sized metropolitan statistical areas ($\sim 1$ million population): Grand Rapids-Wyoming, MI, and Hartford-West Hartford-East Hartford, CT.  
By contrasting and comparing the flow diagrams, we can qualitatively observe that Grand Rapids-Wyoming has a higher frequency of trips from home to work, to community, and to education activities, while Hartford-West Hartford-East Hartford has more trips from home to business, to recreation, and to other destinations.

The trip flow diagram also shows the interconnectedness of different activities.
Focusing on commercial activities, Grand Rapids-Wyoming shows a more even distribution of trips among different O-D pairs compared to Hartford-West Hartford-East Hartford, indicating a more integrated pattern of commercial visits within the overall mobility structure.
On the other hand, community activities in Hartford-West Hartford-East Hartford show a higher degree of interconnectedness than in Grand Rapids-Wyoming.
The ``in-transit'' category also shows differences in the interconnectedness of activities.
Grand Rapids-Wyoming has more trips from ``in-transit'' to ``business'' and ``home'', while Hartford-West Hartford-East Hartford has more trips from ``in-transit'' to ``recreation''.

\subsubsection*{Clustering analysis}

In order to find cities with similar mobility patterns, we cluster the O-D matrices of the $52$ cities in an unsupervised manner. 
Specifically, we use hierarchical clustering, which is widely used because of its high interpretability~\cite{hastie2009elements}, with the Ward method because it tends to find clusters that are balanced and of equal size~\cite{ward1963hierarchical}.


In a few words, Ward's method minimizes the total intra-cluster variance and ensures that the cities grouped together have the most similar trip frequency distributions, as measured by the Euclidean distance between their OD matrices. 
The process starts with each city as a singleton cluster and iteratively merges the pair of clusters that results in the smallest increase in total intra-cluster variance \cite{mullner2011modern, bar2001fast}. 
The result is a dendrogram that visually represents the clustering hierarchy and allows the identification of natural groupings of cities based on their travel patterns. 
This methodology allows us to gain insight into the functional similarities and differences in urban mobility behavior across metropolitan areas.

\begin{table}
\centering\fontsize{10}{10}\selectfont
\begin{tabular}{cc}
\hline
\textbf{2017 NHTS Purpose-Based Categories}                                   & \textbf{Activity-Based Categories} \\ \hline
Reg. Home Activities                                          & Home                     \\
Work from Home (Paid)                                         & Home                     \\
Work                                                          & Work                     \\
Work related/Trip                                             & Work                     \\
Volunteer activities (Not Paid)                               & Community                \\
Drop-off/pickup someone                                       & In Transit               \\
Change type of Transportation                                 & In Transit               \\
Attend school as a student                                    & Education                \\
Attend child care                                             & Care                     \\
Attend adult care                                             & Care                     \\
Buy Goods (groceries, clothes, appliances, gas)               & Shopping                 \\
Buy services (dry cleaners, banking, service a car, pet care) & Shopping                 \\
Buy meals (Go out for a meal, snack, carry-out)               & Meal                 \\
Other general errands (post office, library)                  & Other                    \\
Recreational Activities (visit parks, movies, bars, museums)  & Recreational             \\
Exercise (go for a jog, walk, walk the dog, go to the gym)    & Recreational             \\
Visit Friends and Relatives                                   & Social                   \\
Health care visit (medical, dental, therapy)                  & Care                   \\
Religious or other community activities                       & Community                \\
Something else                                                & Other                    \\ \hline
\end{tabular}

\caption[Activity mappin.]{\textcolor{black}{\textbf{Activity mapping.} 
The $20$ purpose-based categories reported in the $2017$ National Household Travel Survey were mapped into $11$ activity-based categories according to similarity and mobility. For example, this categorization groups the purpose-based categories ``Reg. Home Activities'' and ``Work from Home (Paid)'' into the activity-based one called ``Home,'' reflecting the fact that both purpose-based categories take place at the same location.
}}
\label{table_survey_mapping}
\end{table}

\subsection{Urban form}
\label{sec:urban_form}

\subsubsection*{Input Data}

The NHTS results are reported on an MSA level.
For this reason, we extract MSA boundaries from Census \cite{TIGERLineShapefile2019} and use these boundaries to extract the road networks within cities with OSMnx \cite{boeing2024modeling}, which is a Python library that provides tools for downloading, modeling, analyzing, and visualizing street networks from OpenStreetMap \cite{openstreetmap}.
Once the road networks for the entire MSAs are extracted, we identify the geographic center of the city based on the median latitude and longitude of the nodes of the road networks. 
Then, we select a square box of area $L^2$, with $L = 2$ miles, centered on the geographic center. 
By doing this, we aim to capture the spatial structure and layout of the core area of the cities under consideration \cite{crucitti2006centrality}.
The streets within this core area are turned into an undirected graph $G$ with $N$ nodes and $K$ edges \cite{crucitti2006centrality}. 

\subsubsection*{Knowledge extraction}

After having extracted the road network graph $G$ for each city, we compute four centrality measures for each node $i \in G$: closeness~\cite{freeman2002centrality}, betweenness~\cite{freeman2002centrality}, straightness~\cite{crucitti2006centrality}, and information~\cite{crucitti2006centrality} centralities, which are defined as follows.

\begin{itemize}
    \item \textbf{Closeness centrality}: The closeness centrality \(C_i\) is a measure of how close a node \(i\) is to all other nodes in the network \(G\). It is calculated as the reciprocal of the sum of the shortest path distances from node \(i\) to every other node \(j\) in the network. The formula for closeness centrality is given by:

\begin{equation}
C_i = \frac{N-1}{\sum_{j \in G, j \neq i} d_{ij}},
\end{equation}

\noindent where \(d_{ij}\) is the shortest path distance between nodes \(i\) and \(j\), and \(N\) is the total number of nodes in the network. Higher values of \(C_i\) indicate greater closeness, meaning the node \(i\) is, on average, less distant from all other nodes.

\item \textbf{Betweenness centrality}: Betweenness centrality \(B_i\) quantifies the centrality of a node $i$ by counting the number of shortest paths between each pair of nodes $j$ and $k$ that pass through $i$. 
Nodes with high betweenness centrality are crucial for bridging information traffic across the network, acting as important conduits through which information flows.
This metric is defined as

\begin{equation}
B_i = \sum_{j,k \in G, i \neq j \neq k} \frac{\sigma (j,i,k)}{\sigma (j,k)},
\end{equation}

\noindent where \(\sigma (j,k)\) is the total number of paths from $j$ to $k$ and \(\sigma (j,i,k)\) is the total number of paths from $j$ to $k$  that pass through $i$.

\item \textbf{Straightness centrality}: Straightness centrality \(S_i\) compares the network distance to the Euclidean distance of the nodes. Specifically, for each network node $i$, it measures how much the paths between node $i$ and all other nodes $j$ deviate from a straight line on average. This measure is defined as follows

\begin{equation}
S_i = \frac{1}{N-1} \sum_{j \in G, j \neq i}{\frac{d_{ij}^{\text{Eucl.}}}{d_{ij}}},
\end{equation}

\noindent where \(d_{ij}^{\text{Eucl}}\) is the Euclidean distance between $i$ and $j$.

\item \textbf{Information centrality}: Information centrality \(I_i\) is based on how a network reacts to the deactivation of the node \(i\). Specifically, it is the relative decrease in efficiency when \(i\) is removed from \(G\). This measure is defined as 

\begin{equation}
I_i = \frac{E[G] - E[G'_i]}{E[G]},
\end{equation}

\noindent where \(G'_i\) is obtained from the network \(G\) with all \(N\) nodes and by removing all edges connected to node \(i\), and \(E[G]\) is the efficiency of \(G\), defined as 

\begin{equation}
E[G] = \frac{1}{N(N-1)} \sum_{i,j \in G, i \neq j}{\frac{d_{ij}^{\text{Eucl}}}{d_{ij}}}.
\end{equation}

\end{itemize}


These four measures, $C_i, B_i, S_i, I_i$ provide insights into the accessibility and connectivity of each node $i$ \cite{crucitti2006centrality} of the road network $G$.
This means that each city is represented by four distributions of centrality measures, one for each measure.
Following \cite{crucitti2006centrality}, the heterogeneity of these distributions is captured by the Gini index, which allows us to represent each city by a set of four Gini coefficients: $g^C$ for closeness, $g^B$ for betweenness, $g^S$ for straightness, and $g^I$ for information centrality.
A high Gini coefficient ($g = 1$) indicates significant heterogeneity or inequality within the road network, whereas a low value ($g = 0$) suggests homogeneity.


\subsubsection*{Clustering}

In order to find cities with similar structural patterns, we cluster cities using the same hierarchical clustering method described earlier.
Specifically, each city is represented by $g^C$, $g^B$, $g^S$, and $g^I$.
Then, we use the Ward's method with the Euclidean distance to obtain the hierarchical clusters.
By doing this, we identify coherent clusters of cities that have similar structural patterns as captured by $g^C$, $g^B$, $g^S$, and $g^I$ \cite{crucitti2006centrality}.

\section{Results}
\label{sec:results}


In this study, we cluster 52 U.S. metropolitan areas based on mobility patterns captured by the frequency of trips between the different O-D pairs. Similarly, we cluster the same cities based on their structural similarities.
Capturing ``urban function'', we find 7 clusters of cities where cities in the same cluster exhibit similar travel patterns. At the same time, capturing ``urban form'' we find only two clusters. Notably, there does not appear to be a correspondence between cities clustered together based on function and cities clustered together based on form. The results are described in detail, as follows.



\subsection{Urban function}

\subsubsection*{Clustering based on function}

\begin{figure}[h!]
\centering
\includegraphics[width=1\linewidth]{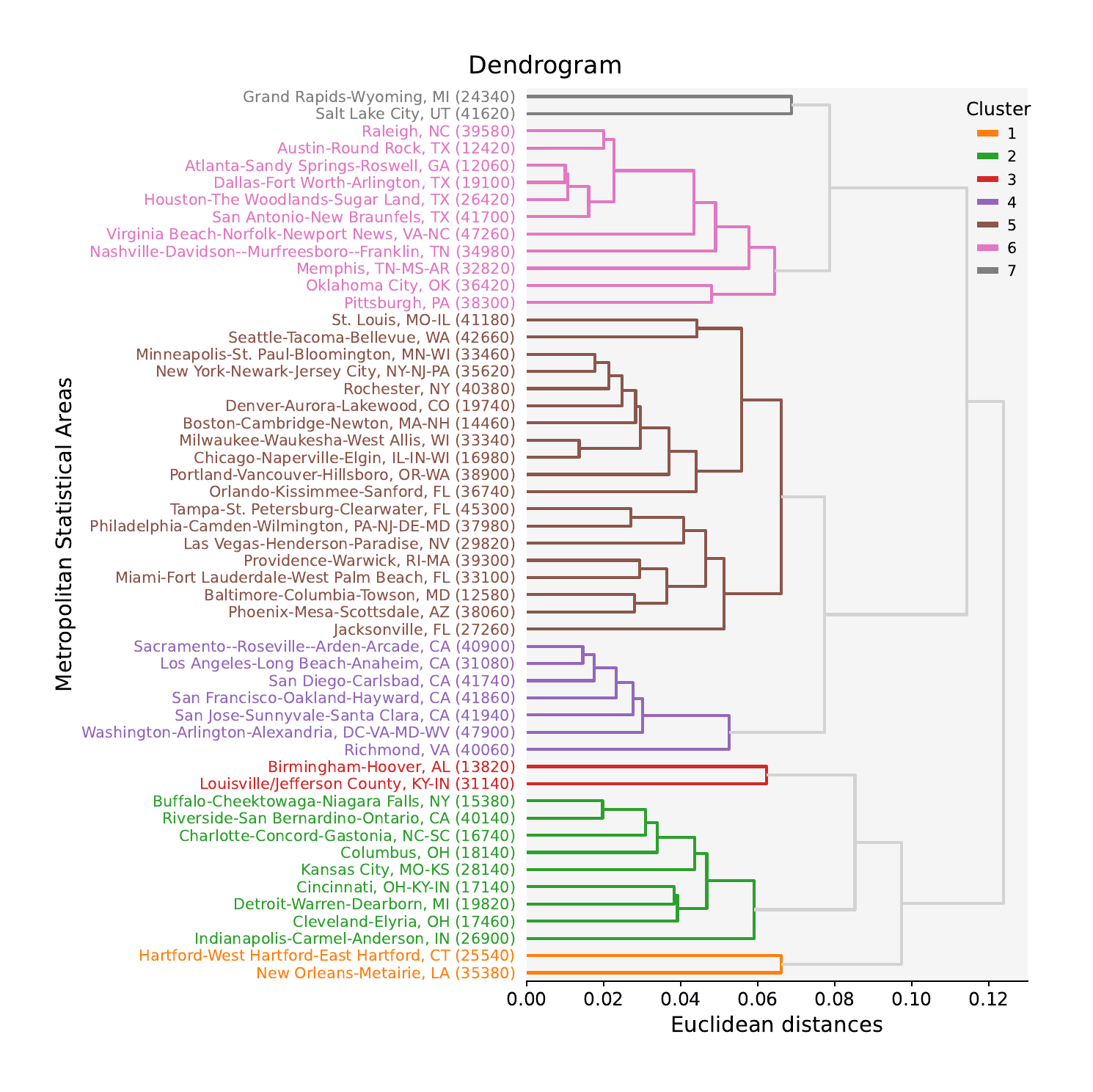}
\caption{\textbf{Clustering metropolitan statistical areas based on the occurrence of trip chains}. 
Hierarchical clustering of cities using flow diagrams, which capture the prevalence of specific trip chains through the frequency of origin-destination (OD) activity pairs. Using a threshold of $60\%$ of the maximum distance, we found that the $52$ metropolitan statistical areas analyzed are grouped into $7$ distinct clusters.}
\label{fig:dendrogram_hierarchical_clustering}
\end{figure}

\begin{figure}[ht]
\centering
\includegraphics[width=1\linewidth, trim={1.5cm 2cm 3cm 2cm},clip]{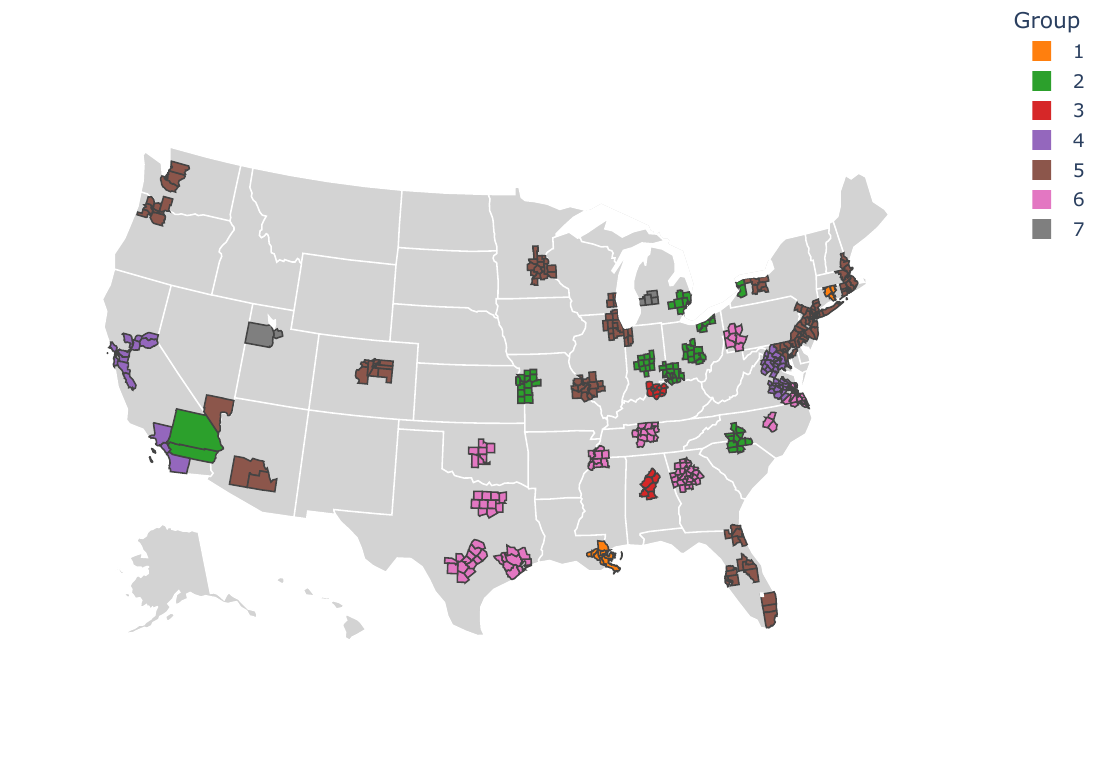}
\caption{
\textbf{Spatial distribution of metropolitan statistical area clusters}.
Boundaries of Metropolitan Statistical Areas at the county level are represented by black lines, and state boundaries are indicated by white lines. 
The color coding of the MSAs corresponds to the colors used in the hierarchical clusters, as shown in Figure \ref{fig:dendrogram_hierarchical_clustering}. 
States such as California, Texas, Florida, and Ohio have multiple Metropolitan Statistical Areas that fall within the same cluster.
}
\label{fig:spatial_distribution_clusters}
\end{figure}

The hierarchical clustering of the $52$ cities under consideration reveals $7$ clusters of cities with similar travel behaviors (Figure \ref{fig:dendrogram_hierarchical_clustering}): three clusters of 2 cities each, and clusters of 7, 9, 11, and 19 cities. 
We observe a tendency for cities within the same U.S. state to be grouped into the same cluster.
For example, all cities in Texas, Wisconsin,  Florida, and Ohio are found in the same cluster, and 5 out of 6 California cities are within the same cluster, suggesting that the local culture of the population somewhat influences mobility patterns. 
Interestingly, however, cities that are geographically distant and have different socioeconomic profiles are also found in the same cluster. 
Consider the cluster consisting of the New Orleans-Metairie and Hartford-West Hartford-East Hartford MSAs, which are separated by more than 1,200 miles and exhibit contrasts in their cultural landscapes, economic drivers, and historical backgrounds.
New Orleans-Metairie, located in the heart of Louisiana's Gulf Coast, has French, Spanish, and African cultural influences. 
The MSA was significantly impacted by Hurricane Katrina in 2005, which resulted in significant population loss and structural damage. 
The economic activity of the city is concentrated in port-related industries, oil and gas extraction, and tourism.
In contrast, Hartford-West Hartford-East Hartford, located in the north-central region of Connecticut, has a more traditional New England character. 
Its economy is driven by the insurance and financial services industries, reflecting its role as a regional commercial center.

The hierarchical clustering also shows that clusters 4 (purple) and 5 (brown) are closely related. 
These clusters consist mainly of cities near the east and west coasts, suggesting similarities in the mobility patterns of coastal residents (Figure \ref{fig:spatial_distribution_clusters}). 
In contrast, the interior regions of the U.S. are mainly composed of cities in clusters 2 (green) and 6 (pink). 
Most cluster 2 cities are located in the northern half of the country, while most cluster 6 cities are located in the southern half.
This spatial distribution may indicate regional differences in mobility behavior, possibly influenced by factors such as climate and cultural practices.

\subsubsection*{Characterizing the clusters}

\begin{figure}[h!]
\centering
\includegraphics{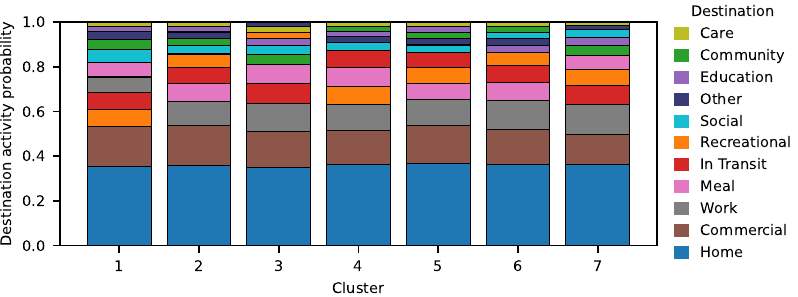}
\caption{
\textbf{Distinct patterns in daily activities and destination priorities across clusters.}
The height of each bar indicates the probability of finding trips to a particular activity, distinguished by different colors. 
Most visited activities are presented at the bottom. 
Note that each cluster has a unique sequence of activities (colors), reflecting the different priorities and preferences in daily activities among the populations of each cluster.
}
\label{fig:stacked_bar_destination_priorities}
\end{figure}

Let us explore the similarities and differences between the clusters of cities.
Figure \ref{fig:stacked_bar_destination_priorities} shows stacked bars illustrating the probability distribution of visits to different activities for each cluster. 
The activities are ordered by visit frequency, with the most visited places at the bottom of each bar and the least visited at the top. 
This arrangement captures the hierarchy of residents' destination preferences, with each activity's position in the stacked bar graph graphically representing its relative importance or priority.

We find that ``home'' is consistently the most visited location across all clusters, accounting for approximately $36\%$ of all visits. This activity is followed by visits to commercial locations in all clusters. 
However, the ranking of other activities shows variations between different clusters. 
For example, ``recreation'' appears as the third most common activity in cities within cluster 1, while ``work'' occupies this position in other clusters. 
Interestingly, in cluster 1, ``work'' ranks fifth, and the visit frequencies for ``recreation'', ``in-transit'', ``work'', ``meal'', and ``social'' are similar, as indicated by the comparable heights of their respective bars in the stacked graph. 
In contrast, cities in the other clusters have a more pronounced difference in the distribution of visits among these activities, with ``work'' being more predominant. 

The observed diversity in activity rankings across city clusters suggests different lifestyle patterns and priorities for activity visits. 
Specifically, in clusters where ``work'' is the third most common activity, there is variation in the fourth most common activity: ``meal'' ranks fourth in clusters 2, 5, and 6; ``recreation'' ranks fourth in cluster 4; ``in-transit'' is the fourth most common activity in clusters 3 and 7. 
In fact, this diversity becomes even more apparent as we move further down the ranking.

\begin{figure}[h!]
\centering
\includegraphics{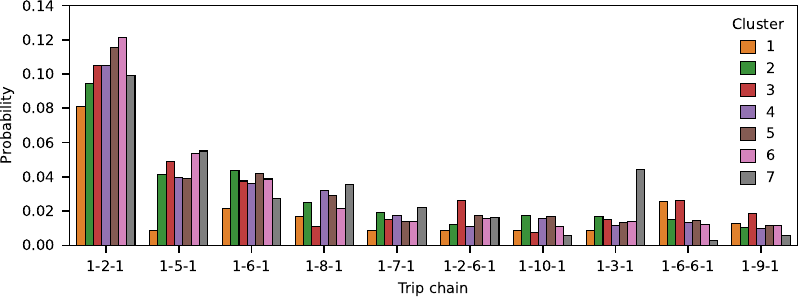}
\caption{
\textbf{Variability in trip chains across city clusters in the U.S.}.
Top 10 most common trip chains nationwide (ranked according to the overall frequency), with activities labeled as: 1 = Home, 2 = Work, 3 = Community, 4 = In Transit, 5 = Education, 6 = Commercial, 7 = Meal, 8 = Recreational, 9 = Care, 10 = Social, 11 = Other. 
Each bar indicates the probability of finding a given activity chain according to the cluster of cities. 
Across all clusters, the ``Home-Work-Home'' chain emerges as the most common. 
However, the variation in probabilities for other sequences highlights the distinct lifestyle patterns in different city clusters.
}
\label{fig:variability_activity_chains}
\end{figure}

\begin{table}[h!]
\centering
\resizebox{\textwidth}{!}{%
\begin{tabular}{cccccccc}
\toprule
\textbf{Rank} & \textbf{Group 1} & \textbf{Group 2} & \textbf{Group 3} & \textbf{Group 4} & \textbf{Group 5} & \textbf{Group 6} & \textbf{Group 7} \\ 
\midrule
\textbf{1st}  & 1-2-1           & 1-2-1            & 1-2-1            & 1-2-1            & 1-2-1            & 1-2-1            & 1-2-1            \\ \hline
\textbf{2nd}  & \makecell{1-4-5-4-1\\ 1-6-6-1}       & 1-6-1            & 1-5-1            & 1-5-1            & 1-6-1            & 1-5-1            & 1-5-1            \\ \hline
\textbf{3rd}  &  1-6-1           & 1-5-1            & 1-6-1            & 1-6-1            & 1-5-1            & 1-6-1            & 1-3-1            \\ \hline
\textbf{4th}  &  \makecell{1-8-1\\ 1-8-6-1\\ 1-10-7-8-10-1}         & 1-8-1            & \makecell{1-2-6-1\\ 1-6-6-1}       & 1-8-1            & 1-8-1            & 1-8-1            & 1-8-1            \\ \hline
\textbf{5th}  & 1-9-1 & 1-7-1 & \makecell{1-6-7-1-6-6-1\\ 1-9-1} & 1-7-1 & 1-2-6-1 & 1-2-6-1 & 1-6-1 \\ 
\bottomrule
\end{tabular}}
\caption{
\textbf{The top 5 trip chains in each city cluster capture the diversity of lifestyle patterns.}
For each group of cities, the table lists the top 5 (most frequent) trip chains, providing insight into the unique behavioral patterns that characterize each cluster.
Activities are labeled as: 1 = Home, 2 = Work, 3 = Community, 4 = In Transit, 5 = Education, 6 = Commercial, 7 = Meal, 8 = Recreational, 9 = Care, 10 = Social, 11 = Other.
}
\label{tab:most_popular_activity_chains_by_cluster}
\end{table}

To better understand the differences in the trip chains of the seven clusters of cities, we show in Figure \ref{fig:variability_activity_chains} the ten most common trip chains in the whole U.S. and the probability of finding each of these chains in the clusters.
We observe that the chain 1-2-1 (home-work-home) is the most frequent in all the clusters, corresponding from $8\%$ (cluster 1) to about $12\%$ (cluster 6) of all the chains.
The chain 1-5-1 (home-education-home) is not very frequent in cluster 1, corresponding to less than $1\%$ of the chains, in contrast to the other clusters where it corresponds to at least $4\%$ of the chains.

While cluster 1 has the lowest frequencies of chain 1-6-1 (home-commercial-home), it has one of the highest frequencies of chain 1-6-6-1 (home-commercial-commercial-home), showing the interconnectedness of commercial activities in the chains of the cities within this cluster.
The same interconnectedness of visits to Commercial places is seen in cities of cluster 3, which have the highest frequencies of chains 1-2-6-1 (home-work-commercial-home) and 1-6-6-1 (home-commercial-commercial-home).
Besides, it is interesting to see the manifestation of religious engagement in cities of cluster 7, where the chain 1-3-1 (home-community-home) is very popular.

Table \ref{tab:most_popular_activity_chains_by_cluster}, which shows the top five most frequent trip chains in each cluster, emphasizes the diversity of trip chains among the clusters. 
The majority of the chains in the table consist of three trips, supporting the idea that most travel patterns are driven by a specific need. 
The 1-2-1 (home-work-home) chain is the most prevalent in all clusters, reflecting the universality of individuals' need to go to Work.
The diversity of trip chains becomes evident from the second most frequent chain onward. 
Cluster 1 cities exhibit longer chains as the second most frequent, indicating a higher degree of interconnectedness between different activities. 
Interestingly, cluster 1 displays the interconnectedness of recreational activities, while cluster 3 exhibits the interconnectedness of commercial activities. 
Again, we observe that each cluster has a unique rank order of the most frequent trip chains.

\subsection{Clustering based on urban form}


The analysis we have presented so far has focused on mobility patterns derived from the frequency of trips and trip chains, regardless of the infrastructure of cities.
Structural aspects of cities, such as road network structure and spatial distribution of different building types, might affect the way people move and schedule their activities \cite{grue2020exploring, naess2012urban, ma2014daily, bautista2020urban}.
In fact, population density might affect the probability of finding complex trip chains \cite{grue2020exploring} since high-density areas imply shorter distances between O-D pairs \cite{naess2012urban}, but car drivers are less likely to be impacted by urban form than users of public transportation \cite{bautista2020urban}.
In the U.S., where about $86\%$ of workers go to work in their own cars \cite{Bloomberg2015DrivingAlone}, and public transportation usage is not only low \cite{anderson2016relies} but also has been decreasing over years \cite{erhardt2022has}, we expect that mobility patterns will not be strongly impacted by urban form.

\begin{figure}[h!]
\centering
\includegraphics[width=1\linewidth]{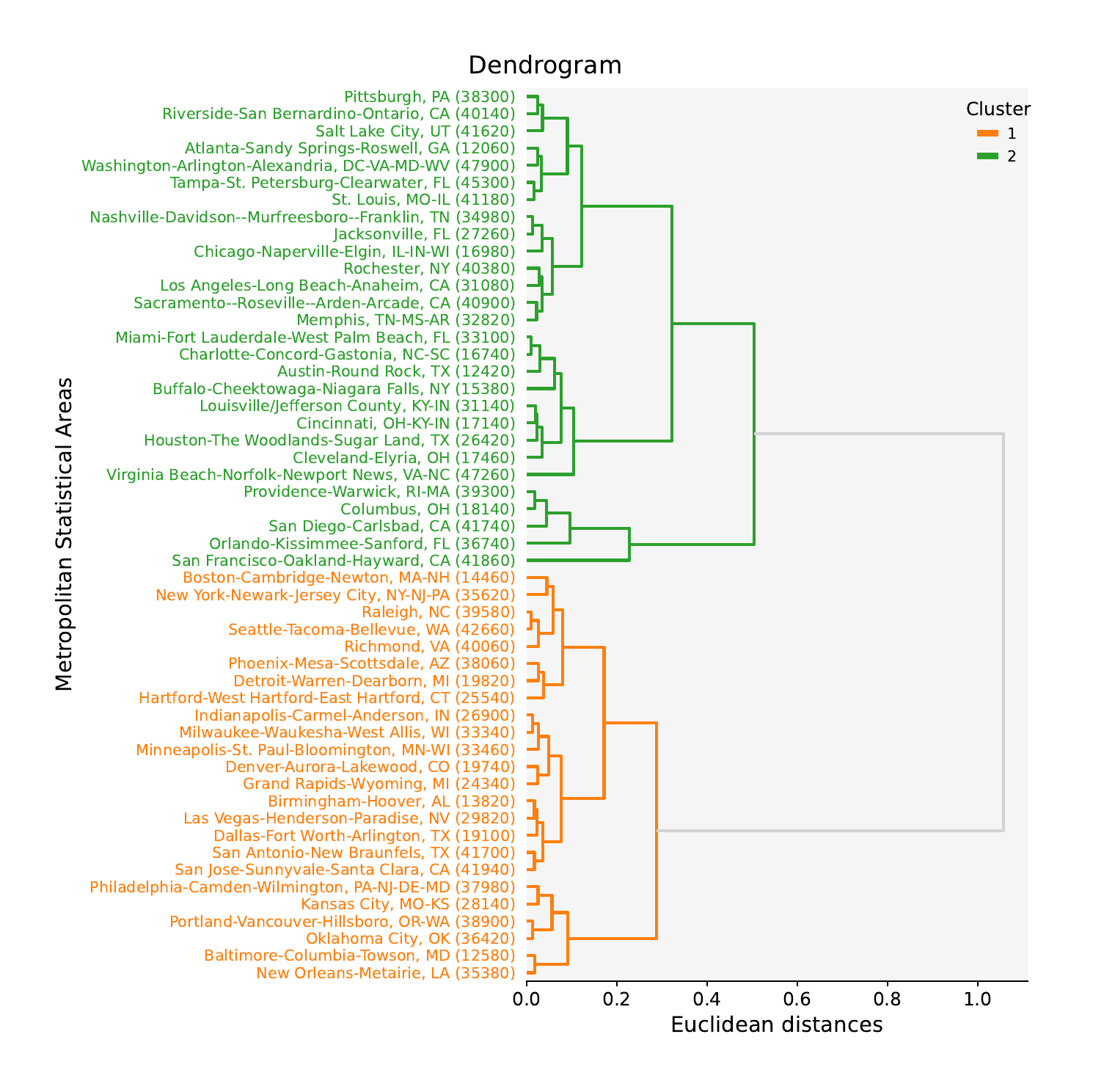}
\caption{\textbf{Clustering metropolitan statistical areas based on the structure of the road networks}. 
Hierarchical clustering of cities using structural features extracted from road networks.
Using a threshold of $60\%$ of the maximum distance, we found that the $52$ metropolitan statistical areas analyzed could be classified into $2$ distinct clusters.}
\label{fig:dendrogram_topological_clustering}
\end{figure}

\begin{figure}[h!]
\centering
\includegraphics{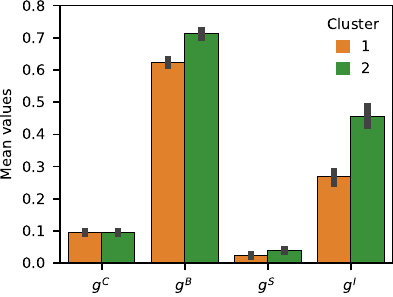}
\caption{
\textbf{Differences between the two structural clusters}.
The Gini coefficients of the centrality metrics used here, $g^C$ for closeness, $g^B$ for betweenness, $g^S$ for straightness, and $g^I$ for information, indicate the differences between the uneven distribution of these metrics among the cities of each cluster.
}
\label{fig:characterizing_topological_clusters}
\end{figure}

In this sense, we also characterize the urban form of the cities under consideration by analyzing four centrality measures (closeness, betweenness, straightness, and information centralities), which are provided in detail in section \ref{sec:urban_form}. 
The pairwise comparison of cities based on their Gini coefficients indicates clusters of cities sharing similar structural properties (Figure \ref{fig:dendrogram_topological_clustering}).
Considering the structural properties, the 52 cities are divided into two clusters. 
Cities in cluster 2 exhibit higher values of $g^B$ and $g^I$ (Figure \ref{fig:characterizing_topological_clusters}), indicating a more heterogeneous structural profile compared to the cities in cluster 1.

In contrast to the clustering by urban function (Figure \ref{fig:dendrogram_hierarchical_clustering}), which shows clear grouping patterns of cities belonging to the same states (e.g., cluster 4 for California and cluster 6 for Texas), the clustering by urban form (Figure \ref{fig:dendrogram_topological_clustering}) does not reveal a strong association between cities that are geographically close.
Although we find pairs of same-state cities, like Dallas--Fort Worth--Arlington and San Antonio--New Braunfels (Texas, cluster 1 in Figure \ref{fig:dendrogram_topological_clustering}), and Los Angeles--Long Beach--Anaheim and Sacramento--Roseville--Arden-Arcade (California, cluster 2 in Figure \ref{fig:dendrogram_topological_clustering}) further down the hierarchy, same-state cities are more scattered across different hierarchical levels.

Moreover, cities that are close to each other in the clustering by function (Figure \ref{fig:dendrogram_hierarchical_clustering}) are found at different hierarchical levels in the urban form clustering (Figure \ref{fig:dendrogram_topological_clustering}). 
For example, in the urban form clustering, the pairs of cities belonging to the three urban function clusters composed of two cities are dispersed. Hartford-West Hartford-East Hartford and New Orleans-Metairie are very close in the function cluster (cluster 1 in Figure \ref{fig:dendrogram_hierarchical_clustering}) but are at different hierarchical levels in the form cluster (Figure \ref{fig:dendrogram_topological_clustering}).
This distinction is even more evident for cities like Birmingham-Hoover and Louisville/Jefferson County, which belong to function cluster 3, and Grand Rapids-Wyoming and Salt Lake City, belonging to function cluster 7. In the urban form clustering, Birmingham-Hoover and Grand Rapids-Wyoming are in form cluster 1, while Louisville/Jefferson County and Salt Lake City are in form cluster 2.
This contrast between the results of the function clustering and the form clustering highlights the dissociation between the form (structural properties) and function (activities) of cities.


\section{Discussion and Conclusions}
\label{sec:concluisons}

The urban-function clustering results suggest that the travel behaviors in U.S. cities can be categorized into a few universal classes. Examining the travel behaviors of cities within clusters reveals similarities and differences that may be tied to the characteristics and needs of populations living in the cities. For example, cities within one cluster have a higher frequency of recreational trips than work trips, which could reflect different socio-demographic characteristics of the population (e.g. younger, retiree) and their needs (e.g. more social activity, more work from home).

The lack of a strong relationship between the urban form and travel behavior indicates that the function of cities may be driven by other mechanisms such as individual needs, socioeconomic factors, and cultural or social dynamics. More research is needed to uncover drivers of these universal mobility classes. We expect infrastructure to have a greater influence on mobility patterns in countries with lower levels of car ownership, where the distance between origin-destination pairs may impose greater constraints on travel patterns.

Future research could benefit from exploring other methods to characterize urban function and urban form. While our study has leveraged specific metrics for urban form and function, there may be other approaches that could reveal alternative insights into the relationship between form and function. Additionally, extending our framework to other versions of the NHTS or travel surveys from different countries could facilitate a broader comparison of urban functions globally. Furthermore, we note that the results are dependent on our aggregation of the trip purpose categories into higher-level activity-based categories and that different results could be obtained using different categories.

In summary, we explored the interplay between urban form and function. Using data from the 2017 NHTS, we propose a framework to characterize the urban function of 52 cities. We also characterize cities with respect to their urban form via centrality measures from road networks. The relationship between form and function can improve urban planning and policy decisions, by basing decisions that drive urban form of a city on the needs and characterizations of the populations that live there.

 \bibliographystyle{elsarticle-num} 
 \bibliography{cas-refs}

\begin{thebibliography}{10}
\expandafter\ifx\csname url\endcsname\relax
  \def\url#1{\texttt{#1}}\fi
\expandafter\ifx\csname urlprefix\endcsname\relax\def\urlprefix{URL }\fi
\expandafter\ifx\csname href\endcsname\relax
  \def\href#1#2{#2} \def\path#1{#1}\fi

\bibitem{allen1997cities}
P.~M. Allen, Cities and regions as evolutionary, complex systems, {Geographical systems} 4 (1997) 103--130.

\bibitem{bettencourt2015cities}
L.~M. Bettencourt, Cities as complex systems, Modeling complex systems for public policies (2015) 217--236.

\bibitem{batty1989urban}
M.~Batty, P.~Longley, S.~Fotheringham, {Urban growth and form: scaling, fractal geometry, and diffusion-limited aggregation}, {Environment and planning A} 21~(11) (1989) 1447--1472.

\bibitem{batty1991cities}
M.~Batty, {Cities as fractals: simulating growth and form}, in: Fractals and chaos, Springer, 1991, pp. 43--69.

\bibitem{frankhauser2004comparing}
P.~Frankhauser, {Comparing the morphology of urban patterns in Europe--a fractal approach}, {European Cities--Insights on outskirts, Report COST Action} 10 (2004) 79--105.

\bibitem{verbavatz2020growth}
V.~Verbavatz, M.~Barthelemy, {The growth equation of cities}, {Nature} 587~(7834) (2020) 397--401.

\bibitem{bettencourt2020demography}
L.~M. Bettencourt, D.~Z{\"u}nd, Demography and the emergence of universal patterns in urban systems, Nature communications 11~(1) (2020) 4584.

\bibitem{reia2022spatial}
S.~M. Reia, P.~S.~C. Rao, M.~Barthelemy, S.~V. Ukkusuri, Spatial structure of city population growth, Nature communications 13~(1) (2022) 5931.

\bibitem{reia2022modeling}
S.~M. Reia, P.~S.~C. Rao, S.~V. Ukkusuri, Modeling the dynamics and spatial heterogeneity of city growth, {npj Urban Sustainability} 2~(1) (2022) 31.

\bibitem{primerano2008defining}
F.~Primerano, M.~A. Taylor, L.~Pitaksringkarn, P.~Tisato, {Defining and understanding trip chaining behaviour}, Transportation 35 (2008) 55--72.

\bibitem{hu2021modeling}
Y.~Hu, X.~Li, Modeling and analysis of excess commuting with trip chains, Annals of the American Association of Geographers 111~(6) (2021) 1851--1867.

\bibitem{clifton2008quantitative}
K.~Clifton, R.~Ewing, G.-J. Knaap, Y.~Song, Quantitative analysis of urban form: a multidisciplinary review, Journal of Urbanism 1~(1) (2008) 17--45.

\bibitem{crooks2015crowdsourcing}
A.~Crooks, D.~Pfoser, A.~Jenkins, A.~Croitoru, A.~Stefanidis, D.~Smith, S.~Karagiorgou, A.~Efentakis, G.~Lamprianidis, Crowdsourcing urban form and function, {International Journal of Geographical Information Science} 29~(5) (2015) 720--741.

\bibitem{dascher2019function}
K.~Dascher, Function follows form, Journal of Housing Economics 44 (2019) 131--140.

\bibitem{batty1992form}
M.~Batty, K.~Sik~Kim, Form follows function: reformulating urban population density functions, Urban studies 29~(7) (1992) 1043--1069.

\bibitem{duany2000suburban}
A.~Duany, E.~Plater-Zyberk, J.~Speck, Suburban nation: The rise of sprawl and the decline of the American dream, Macmillan, 2000.

\bibitem{burger2012form}
M.~Burger, E.~Meijers, Form follows function? linking morphological and functional polycentricity, Urban studies 49~(5) (2012) 1127--1149.

\bibitem{van2011mapping}
T.~Van~de Voorde, W.~Jacquet, F.~Canters, {Mapping form and function in urban areas: An approach based on urban metrics and continuous impervious surface data}, {Landscape and Urban Planning} 102~(3) (2011) 143--155.

\bibitem{FHWA2022}
{Federal Highway Administration}, \href{http://nhts.ornl.gov}{{2022 NextGen National Household Travel Survey Core Data}}, U.S. Department of Transportation (2022).
\newline\urlprefix\url{http://nhts.ornl.gov}

\bibitem{crucitti2006centrality}
P.~Crucitti, V.~Latora, S.~Porta, Centrality measures in spatial networks of urban streets, {Physical Review E} 73~(3) (2006) 036125.

\bibitem{maslow1987maslow}
A.~Maslow, K.~Lewis, Maslow's hierarchy of needs, {Salenger Incorporated} 14~(17) (1987) 987--990.

\bibitem{boccaletti2006complex}
S.~Boccaletti, V.~Latora, Y.~Moreno, M.~Chavez, D.-U. Hwang, Complex networks: Structure and dynamics, Physics reports 424~(4-5) (2006) 175--308.

\bibitem{costa2007characterization}
L.~da~F~Costa, F.~A. Rodrigues, G.~Travieso, P.~R. Villas~Boas, Characterization of complex networks: A survey of measurements, Advances in physics 56~(1) (2007) 167--242.

\bibitem{alexander2015origin}
L.~Alexander, S.~Jiang, M.~Murga, M.~C. Gonz{\'a}lez, {Origin--destination trips by purpose and time of day inferred from mobile phone data}, {Transportation research part c: emerging technologies} 58 (2015) 240--250.

\bibitem{wang2018applying}
Z.~Wang, S.~Y. He, Y.~Leung, {Applying mobile phone data to travel behaviour research: A literature review}, {Travel Behaviour and Society} 11 (2018) 141--155.

\bibitem{holguin2005observed}
J.~Holgu{\'\i}n-Veras, G.~R. Patil, Observed trip chain behavior of commercial vehicles, {Transportation research record} 1906~(1) (2005) 74--80.

\bibitem{mcguckin2004trips}
N.~McGuckin, Y.~Nakamoto, {Trips, Chains and tours-using an operational definition}, in: {National Household Travel Survey Conference}, 2004, p.~55.

\bibitem{chauhan2021database}
R.~S. Chauhan, M.~W. Bhagat-Conway, D.~Capasso~da Silva, D.~Salon, A.~Shamshiripour, E.~Rahimi, S.~Khoeini, A.~Mohammadian, S.~Derrible, R.~Pendyala, {A database of travel-related behaviors and attitudes before, during, and after COVID-19 in the United States}, {Scientific Data} 8~(1) (2021) 245.

\bibitem{jiang2013review}
S.~Jiang, G.~A. Fiore, Y.~Yang, J.~Ferreira~Jr, E.~Frazzoli, M.~C. Gonz{\'a}lez, {A review of urban computing for mobile phone traces: current methods, challenges and opportunities}, in: {Proceedings of the 2nd ACM SIGKDD international workshop on Urban Computing}, 2013, pp. 1--9.

\bibitem{wang2017trip}
L.~Wang, W.~Ma, Y.~Fan, Z.~Zuo, {Trip chain extraction using smartphone-collected trajectory data}, {Transportmetrica B: Transport Dynamics} (2017).

\bibitem{safi2017empirical}
H.~Safi, B.~Assemi, M.~Mesbah, L.~Ferreira, An empirical comparison of four technology-mediated travel survey methods, Journal of traffic and transportation engineering (English edition) 4~(1) (2017) 80--87.

\bibitem{schafer2000regularities}
A.~Schafer, Regularities in travel demand: an international perspective (2000).

\bibitem{stopher2011national}
P.~Stopher, Y.~Zhang, J.~Armoogum, J.-L. Madre, National household travel surveys: The case for australia, in: 34th Australasian Transport Research Forum (ATRF), Adelaide, South Australia, Citeseer, 2011.

\bibitem{NTTS_2009}
{Data and Statistical Studies Service (SdES)}, \href{https://www.insee.fr/en/metadonnees/source/serie/s1277}{{National transport and travel survey}}, InSee (2009).
\newline\urlprefix\url{https://www.insee.fr/en/metadonnees/source/serie/s1277}

\bibitem{NTS_2023}
{Department for Transport}, \href{DOI: http://doi.org/10.5255/UKDA-Series-2000037}{{National Travel Survey}}, UK Data Service (2009).
\newline\urlprefix\url{DOI: http://doi.org/10.5255/UKDA-Series-2000037}

\bibitem{cbs_dutch_national_travel_survey}
{Statistics Netherlands (CBS)}, \href{https://www.cbs.nl/en-gb/our-services/methods/surveys/brief-survey-description/dutch-national-travel-survey}{Dutch national travel survey}, accessed: 2024-06-05 (2023).
\newline\urlprefix\url{https://www.cbs.nl/en-gb/our-services/methods/surveys/brief-survey-description/dutch-national-travel-survey}

\bibitem{ssb_norwegian_travel_survey}
{Statistics Norway}, \href{https://www.ssb.no/en/transport-og-reiseliv/reiseliv/statistikk/reiseundersokelsen}{Norwegian travel survey (reiseundersøkelsen)}, accessed: 2024-06-05 (2023).
\newline\urlprefix\url{https://www.ssb.no/en/transport-og-reiseliv/reiseliv/statistikk/reiseundersokelsen}

\bibitem{bfs_swiss_travel_survey}
{Federal Statistical Office (FSO), Switzerland}, \href{https://www.bfs.admin.ch/bfs/en/home/statistics/tourism/surveys/rv.html}{Swiss travel survey (reiseverhalten)}, accessed: 2024-06-05 (2023).
\newline\urlprefix\url{https://www.bfs.admin.ch/bfs/en/home/statistics/tourism/surveys/rv.html}

\bibitem{statistik_at_austrian_travel_habits_survey}
{Statistics Austria}, \href{https://www.statistik.at/en/ueber-uns/erhebungen/personen-und-haushaltserhebungen/travel-habits}{Travel habits survey}, accessed: 2024-06-05 (2023).
\newline\urlprefix\url{https://www.statistik.at/en/ueber-uns/erhebungen/personen-und-haushaltserhebungen/travel-habits}

\bibitem{statcan_national_travel_survey}
{Statistics Canada}, \href{https://www.statcan.gc.ca/en/survey/household/5232}{National travel survey}, accessed: 2024-06-05 (2023).
\newline\urlprefix\url{https://www.statcan.gc.ca/en/survey/household/5232}

\bibitem{gonzalez2008understanding}
M.~C. Gonzalez, C.~A. Hidalgo, A.-L. Barabasi, Understanding individual human mobility patterns, Nature 453~(7196) (2008) 779--782.

\bibitem{schneider2013unravelling}
C.~M. Schneider, V.~Belik, T.~Couronn{\'e}, Z.~Smoreda, M.~C. Gonz{\'a}lez, Unravelling daily human mobility motifs, {Journal of The Royal Society Interface} 10~(84) (2013) 20130246.

\bibitem{ectors2019exploratory}
W.~Ectors, B.~Kochan, D.~Janssens, T.~Bellemans, G.~Wets, {Exploratory analysis of Zipf’s universal power law in activity schedules}, {Transportation} 46~(5) (2019) 1689--1712.

\bibitem{vzivkovic2020urban}
J.~{\v{Z}}ivkovi{\'c}, Urban form and function, Climate action (2020) 862--871.

\bibitem{kropf2009aspects}
K.~Kropf, Aspects of urban form, Urban morphology 13~(2) (2009) 105--120.

\bibitem{hillier1989social}
B.~Hillier, J.~Hanson, The social logic of space, Cambridge university press, 1989.

\bibitem{jiang2004topological}
B.~Jiang, C.~Claramunt, Topological analysis of urban street networks, {Environment and Planning B: Planning and design} 31~(1) (2004) 151--162.

\bibitem{porta2006network}
S.~Porta, P.~Crucitti, V.~Latora, The network analysis of urban streets: A dual approach, {Physica A: Statistical Mechanics and its Applications} 369~(2) (2006) 853--866.

\bibitem{mesev1995morphology}
T.~Mesev, P.~A. Longley, M.~Batty, Y.~Xie, Morphology from imagery: detecting and measuring the density of urban land use, {Environment and planning A} 27~(5) (1995) 759--780.

\bibitem{herold2003spatial}
M.~Herold, X.~Liu, K.~C. Clarke, Spatial metrics and image texture for mapping urban land use, {Photogrammetric Engineering \& Remote Sensing} 69~(9) (2003) 991--1001.

\bibitem{wang2023eo+}
J.~Wang, M.~Fleischmann, A.~Venerandi, O.~Romice, M.~Kuffer, S.~Porta, Eo+ morphometrics: Understanding cities through urban morphology at large scale, {Landscape and Urban Planning} 233 (2023) 104691.

\bibitem{lynch1964image}
K.~Lynch, The image of the city, MIT press, 1964.

\bibitem{jacobsdeath}
J.~Jacobs, The death and life of great american cities; reissue edition; vintage: New york, ny, usa, 1992, Google Scholar.

\bibitem{louf2013modeling}
R.~Louf, M.~Barthelemy, Modeling the polycentric transition of cities, Physical review letters 111~(19) (2013) 198702.

\bibitem{crooks2015agent}
A.~Crooks, N.~Malleson, E.~Manley, A.~Heppenstall, Agent-based modeling and geographical information systems, {Geocomputation: A Practical Primer. SAGE Publications Ltd, Thousand Oaks, CA} (2015) 63--77.

\bibitem{balac2021synthetic}
M.~Bala{\'c}, S.~H{\"o}rl, {Synthetic population for the state of California based on open data: Examples of the San Francisco Bay Area and San Diego County}, in: {100th Annual Meeting of the Transportation Research Board (TRB 2021)}, IVT, ETH Zurich, 2021.

\bibitem{hastie2009elements}
T.~Hastie, R.~Tibshirani, J.~H. Friedman, J.~H. Friedman, The elements of statistical learning: data mining, inference, and prediction, Vol.~2, Springer, 2009.

\bibitem{ward1963hierarchical}
J.~H. Ward~Jr, Hierarchical grouping to optimize an objective function, Journal of the American statistical association 58~(301) (1963) 236--244.

\bibitem{mullner2011modern}
D.~M{\"u}llner, Modern hierarchical, agglomerative clustering algorithms, arXiv preprint arXiv:1109.2378 (2011).

\bibitem{bar2001fast}
Z.~Bar-Joseph, D.~K. Gifford, T.~S. Jaakkola, Fast optimal leaf ordering for hierarchical clustering, Bioinformatics 17~(suppl\_1) (2001) S22--S29.

\bibitem{TIGERLineShapefile2019}
{U.S. Census Bureau}, {TIGER/Line Shapefile, 2019, nation, U.S., Current Metropolitan Statistical Area/Micropolitan Statistical Area (CBSA) National}, \url{https://catalog.data.gov/dataset/tiger-line-shapefile-2019-nation-u-s-current-metropolitan-statistical-area-micropolitan-statist}, accessed: 2024-04-03 (2019).

\bibitem{boeing2024modeling}
G.~Boeing, Modeling and analyzing urban networks and amenities with osmnx (2024).

\bibitem{openstreetmap}
{OpenStreetMap contributors}, \href{https://www.openstreetmap.org}{Openstreetmap}, accessed: 2024-05-14 (2024).
\newline\urlprefix\url{https://www.openstreetmap.org}

\bibitem{freeman2002centrality}
L.~C. Freeman, et~al., Centrality in social networks: Conceptual clarification, Social network: critical concepts in sociology. Londres: Routledge 1 (2002) 238--263.

\bibitem{grue2020exploring}
B.~Grue, K.~Veisten, {\O}.~Engebretsen, {Exploring the relationship between the built environment, trip chain complexity, and auto mode choice, applying a large national data set}, {Transportation Research Interdisciplinary Perspectives} 5 (2020) 100134.

\bibitem{naess2012urban}
P.~N{\ae}ss, {Urban form and travel behavior: Experience from a Nordic context}, {Journal of Transport and Land use} 5~(2) (2012) 21--45.

\bibitem{ma2014daily}
J.~Ma, G.~Mitchell, A.~Heppenstall, {Daily travel behaviour in Beijing, China: An analysis of workers' trip chains, and the role of socio-demographics and urban form}, {Habitat International} 43 (2014) 263--273.

\bibitem{bautista2020urban}
D.~Bautista-Hern{\'a}ndez, {Urban structure and its influence on trip chaining complexity in the Mexico City Metropolitan Area}, {Urban, Planning and Transport Research} 8~(1) (2020) 71--97.

\bibitem{Bloomberg2015DrivingAlone}
\href{https://www.bloomberg.com/news/articles/2015-08-17/america-s-ongoing-love-affair-with-the-driving-to-work-alone}{{America's Ongoing Love Affair With the Car}} (Aug 2015).
\newline\urlprefix\url{https://www.bloomberg.com/news/articles/2015-08-17/america-s-ongoing-love-affair-with-the-driving-to-work-alone}

\bibitem{anderson2016relies}
M.~Anderson, {Who relies on public transit in the US}, {Pew Research Center} 7 (2016).

\bibitem{erhardt2022has}
G.~D. Erhardt, J.~M. Hoque, V.~Goyal, S.~Berrebi, C.~Brakewood, K.~E. Watkins, {Why has public transit ridership declined in the United States?}, {Transportation research part A: policy and practice} 161 (2022) 68--87.

\end{thebibliography}

\appendix
\begin{table}
\centering

\resizebox{\textwidth}{!}{\begin{tabular}{llrrrrrr}
\toprule
CBSA & Name & Population & Trips & Households & Respondents & $R_P$ & $T_R$ \\
\midrule
35620 & New York-Newark-Jersey City, NY-NJ-PA & 19995910 & 40600 & 5537 & 11759 & 0.000588 & 3.452675 \\
31080 & Los Angeles-Long Beach-Anaheim, CA & 13278000 & 22947 & 3178 & 6627 & 0.000499 & 3.462653 \\
16980 & Chicago-Naperville-Elgin, IL-IN-WI & 9514113 & 6955 & 909 & 1890 & 0.000199 & 3.679894 \\
19100 & Dallas-Fort Worth-Arlington, TX & 7403925 & 66565 & 8988 & 19016 & 0.002568 & 3.500473 \\
26420 & Houston-The Woodlands-Sugar Land, TX & 6900090 & 35171 & 4803 & 10302 & 0.001493 & 3.413997 \\
47900 & Washington-Arlington-Alexandria, DC-VA-MD-WV & 6198129 & 6001 & 843 & 1687 & 0.000272 & 3.557202 \\
33100 & Miami-Fort Lauderdale-West Palm Beach, FL & 6118155 & 1986 & 305 & 590 & 0.000096 & 3.366102 \\
37980 & Philadelphia-Camden-Wilmington, PA-NJ-DE-MD & 6078522 & 4420 & 651 & 1273 & 0.000209 & 3.472113 \\
12060 & Atlanta-Sandy Springs-Roswell, GA & 5872432 & 19682 & 2791 & 5790 & 0.000986 & 3.399309 \\
14460 & Boston-Cambridge-Newton, MA-NH & 4841772 & 3030 & 398 & 804 & 0.000166 & 3.768657 \\
38060 & Phoenix-Mesa-Scottsdale, AZ & 4758748 & 4875 & 735 & 1465 & 0.000308 & 3.327645 \\
41860 & San Francisco-Oakland-Hayward, CA & 4712421 & 17609 & 2309 & 4688 & 0.000995 & 3.756186 \\
40140 & Riverside-San Bernardino-Ontario, CA & 4565909 & 6519 & 968 & 2128 & 0.000466 & 3.063440 \\
19820 & Detroit-Warren-Dearborn, MI & 4321593 & 2169 & 308 & 618 & 0.000143 & 3.509709 \\
42660 & Seattle-Tacoma-Bellevue, WA & 3885579 & 2393 & 343 & 702 & 0.000181 & 3.408832 \\
33460 & Minneapolis-St. Paul-Bloomington, MN-WI & 3590598 & 4734 & 656 & 1354 & 0.000377 & 3.496307 \\
41740 & San Diego-Carlsbad, CA & 3321237 & 20375 & 2775 & 5674 & 0.001708 & 3.590941 \\
45300 & Tampa-St. Petersburg-Clearwater, FL & 3106922 & 1574 & 252 & 461 & 0.000148 & 3.414317 \\
19740 & Denver-Aurora-Lakewood, CO & 2891776 & 2033 & 256 & 519 & 0.000179 & 3.917148 \\
41180 & St. Louis, MO-IL & 2805758 & 1849 & 240 & 502 & 0.000179 & 3.683267 \\
12580 & Baltimore-Columbia-Towson, MD & 2798707 & 2793 & 411 & 834 & 0.000298 & 3.348921 \\
16740 & Charlotte-Concord-Gastonia, NC-SC & 2525544 & 6168 & 845 & 1705 & 0.000675 & 3.617595 \\
36740 & Orlando-Kissimmee-Sanford, FL & 2517777 & 1151 & 153 & 331 & 0.000131 & 3.477341 \\
41700 & San Antonio-New Braunfels, TX & 2472121 & 13672 & 1897 & 3959 & 0.001601 & 3.453397 \\
38900 & Portland-Vancouver-Hillsboro, OR-WA & 2454815 & 1836 & 228 & 493 & 0.000201 & 3.724138 \\
38300 & Pittsburgh, PA & 2329004 & 1511 & 223 & 435 & 0.000187 & 3.473563 \\
40900 & Sacramento--Roseville--Arden-Arcade, CA & 2319572 & 28027 & 3984 & 8287 & 0.003573 & 3.382044 \\
29820 & Las Vegas-Henderson-Paradise, NV & 2181635 & 1142 & 158 & 317 & 0.000145 & 3.602524 \\
17140 & Cincinnati, OH-KY-IN & 2179864 & 1326 & 172 & 357 & 0.000164 & 3.714286 \\
28140 & Kansas City, MO-KS & 2127203 & 1228 & 163 & 321 & 0.000151 & 3.825545 \\
12420 & Austin-Round Rock, TX & 2115475 & 16205 & 2168 & 4441 & 0.002099 & 3.648953 \\
18140 & Columbus, OH & 2082581 & 1461 & 188 & 377 & 0.000181 & 3.875332 \\
17460 & Cleveland-Elyria, OH & 2057238 & 1409 & 187 & 382 & 0.000186 & 3.688482 \\
26900 & Indianapolis-Carmel-Anderson, IN & 2027584 & 1052 & 152 & 297 & 0.000146 & 3.542088 \\
41940 & San Jose-Sunnyvale-Santa Clara, CA & 1992674 & 7504 & 939 & 2084 & 0.001046 & 3.600768 \\
34980 & Nashville-Davidson--Murfreesboro--Franklin, TN & 1899354 & 837 & 126 & 256 & 0.000135 & 3.269531 \\
47260 & Virginia Beach-Norfolk-Newport News, VA-NC & 1724408 & 1364 & 190 & 407 & 0.000236 & 3.351351 \\
39300 & Providence-Warwick, RI-MA & 1616614 & 1844 & 275 & 533 & 0.000330 & 3.459662 \\
33340 & Milwaukee-Waukesha-West Allis, WI & 1574444 & 21800 & 2913 & 5787 & 0.003676 & 3.767064 \\
27260 & Jacksonville, FL & 1505033 & 765 & 111 & 215 & 0.000143 & 3.558140 \\
36420 & Oklahoma City, OK & 1381492 & 760 & 105 & 210 & 0.000152 & 3.619048 \\
32820 & Memphis, TN-MS-AR & 1346837 & 544 & 74 & 160 & 0.000119 & 3.400000 \\
39580 & Raleigh, NC & 1334235 & 4437 & 561 & 1203 & 0.000902 & 3.688279 \\
40060 & Richmond, VA & 1292999 & 803 & 111 & 211 & 0.000163 & 3.805687 \\
31140 & Louisville/Jefferson County, KY-IN & 1291867 & 671 & 102 & 191 & 0.000148 & 3.513089 \\
35380 & New Orleans-Metairie, LA & 1270326 & 570 & 76 & 140 & 0.000110 & 4.071429 \\
25540 & Hartford-West Hartford-East Hartford, CT & 1207027 & 710 & 94 & 187 & 0.000155 & 3.796791 \\
41620 & Salt Lake City, UT & 1204205 & 1068 & 124 & 294 & 0.000244 & 3.632653 \\
13820 & Birmingham-Hoover, AL & 1149510 & 604 & 80 & 174 & 0.000151 & 3.471264 \\
15380 & Buffalo-Cheektowaga-Niagara Falls, NY & 1129882 & 6076 & 845 & 1664 & 0.001473 & 3.651442 \\
40380 & Rochester, NY & 1071962 & 6748 & 963 & 1945 & 0.001814 & 3.469409 \\
24340 & Grand Rapids-Wyoming, MI & 1060068 & 706 & 87 & 182 & 0.000172 & 3.879121 \\
\bottomrule
\end{tabular}}

\caption[Survey sample of the $52$ metropolitan statistical areas considered in our study.]{\textcolor{black}{\textbf{Survey sample of the $52$ metropolitan statistical areas considered in our study.} 
The populations of the $52$ MSAs range from about $20$ million (New York-Newark-Jersey City, NY-NJ-PA) to about $1$ million (Grand Rapids-Wyoming, MI).
Miami-Fort Lauderdale-West Palm Beach, FL is the MSA with the lowest number of respondents per population ($R_P$), while Milwaukee-Waukesha-West Allis, WI is the MSA with the highest $R_P$.
New Orleans-Metairie, LA is the MSA with the highest number of trips per respondent ($T_R$), while Riverside-San Bernardino-Ontario, CA is the MSA with the lowest $T_R$.
}}
\label{table_survey_sample}
\end{table}

\begin{table}
\centering

\resizebox{\textwidth}{!}{\begin{tabular}{rlrrrrrr}
\toprule
CBSA & Name & N & K & $g^C$ & $g^B$ & $g^S$ & $g^I$ \\
\midrule
35620 & New York-Newark-Jersey City, NY-NJ-PA & 2394 & 5387 & 0.127327 & 0.682789 & 0.039603 & 0.318825 \\
31080 & Los Angeles-Long Beach-Anaheim, CA & 2152 & 5441 & 0.094496 & 0.709616 & 0.035135 & 0.412317 \\
16980 & Chicago-Naperville-Elgin, IL-IN-WI & 1128 & 2647 & 0.073475 & 0.685659 & 0.041714 & 0.458221 \\
19100 & Dallas-Fort Worth-Arlington, TX & 1207 & 2422 & 0.080747 & 0.615434 & 0.035558 & 0.271236 \\
26420 & Houston-The Woodlands-Sugar Land, TX & 1919 & 4488 & 0.082815 & 0.691538 & 0.036149 & 0.373662 \\
47900 & Washington-Arlington-Alexandria, DC-VA-MD-WV & 2486 & 5815 & 0.087535 & 0.725811 & 0.043323 & 0.475085 \\
33100 & Miami-Fort Lauderdale-West Palm Beach, FL & 2096 & 5229 & 0.088166 & 0.676180 & 0.027919 & 0.348319 \\
37980 & Philadelphia-Camden-Wilmington, PA-NJ-DE-MD & 3365 & 8734 & 0.110400 & 0.590990 & 0.018681 & 0.220553 \\
12060 & Atlanta-Sandy Springs-Roswell, GA & 1922 & 4491 & 0.097634 & 0.730911 & 0.036503 & 0.495538 \\
14460 & Boston-Cambridge-Newton, MA-NH & 4614 & 10601 & 0.086550 & 0.686178 & 0.021993 & 0.316404 \\
38060 & Phoenix-Mesa-Scottsdale, AZ & 2087 & 5505 & 0.086680 & 0.630859 & 0.021133 & 0.333990 \\
41860 & San Francisco-Oakland-Hayward, CA & 296 & 612 & 0.120070 & 0.785144 & 0.073210 & 0.730222 \\
40140 & Riverside-San Bernardino-Ontario, CA & 1235 & 2886 & 0.098547 & 0.755147 & 0.041215 & 0.454126 \\
19820 & Detroit-Warren-Dearborn, MI & 2138 & 6059 & 0.077572 & 0.633654 & 0.024718 & 0.310755 \\
42660 & Seattle-Tacoma-Bellevue, WA & 1770 & 4343 & 0.106320 & 0.654941 & 0.028030 & 0.300125 \\
33460 & Minneapolis-St. Paul-Bloomington, MN-WI & 2404 & 6546 & 0.103187 & 0.626181 & 0.024254 & 0.264981 \\
41740 & San Diego-Carlsbad, CA & 534 & 1185 & 0.119378 & 0.733981 & 0.042088 & 0.531494 \\
45300 & Tampa-St. Petersburg-Clearwater, FL & 1917 & 4503 & 0.087433 & 0.704628 & 0.041332 & 0.482138 \\
19740 & Denver-Aurora-Lakewood, CO & 2365 & 7132 & 0.093128 & 0.582267 & 0.020100 & 0.250526 \\
41180 & St. Louis, MO-IL & 2183 & 5441 & 0.097427 & 0.706505 & 0.032111 & 0.491463 \\
12580 & Baltimore-Columbia-Towson, MD & 3537 & 8981 & 0.097876 & 0.588500 & 0.014014 & 0.165263 \\
16740 & Charlotte-Concord-Gastonia, NC-SC & 2378 & 5992 & 0.097786 & 0.676392 & 0.028119 & 0.344950 \\
36740 & Orlando-Kissimmee-Sanford, FL & 1643 & 4085 & 0.106687 & 0.679974 & 0.051443 & 0.595827 \\
41700 & San Antonio-New Braunfels, TX & 1791 & 4714 & 0.081657 & 0.621472 & 0.023974 & 0.299523 \\
38900 & Portland-Vancouver-Hillsboro, OR-WA & 4435 & 12416 & 0.099747 & 0.626652 & 0.018926 & 0.220801 \\
38300 & Pittsburgh, PA & 2177 & 5795 & 0.099360 & 0.751551 & 0.047649 & 0.477508 \\
40900 & Sacramento--Roseville--Arden-Arcade, CA & 2768 & 6664 & 0.096988 & 0.724939 & 0.036631 & 0.429907 \\
29820 & Las Vegas-Henderson-Paradise, NV & 2338 & 5360 & 0.081965 & 0.625770 & 0.027804 & 0.282615 \\
17140 & Cincinnati, OH-KY-IN & 1738 & 4574 & 0.095197 & 0.687736 & 0.028983 & 0.389251 \\
28140 & Kansas City, MO-KS & 2503 & 7452 & 0.097335 & 0.575030 & 0.011462 & 0.207925 \\
12420 & Austin-Round Rock, TX & 1943 & 4985 & 0.082082 & 0.662383 & 0.030088 & 0.365086 \\
18140 & Columbus, OH & 1397 & 3533 & 0.096888 & 0.746490 & 0.051459 & 0.562358 \\
17460 & Cleveland-Elyria, OH & 1564 & 4336 & 0.091482 & 0.671150 & 0.030674 & 0.400152 \\
26900 & Indianapolis-Carmel-Anderson, IN & 2073 & 6071 & 0.089505 & 0.625797 & 0.019112 & 0.251671 \\
41940 & San Jose-Sunnyvale-Santa Clara, CA & 2676 & 6730 & 0.094453 & 0.610913 & 0.021256 & 0.302443 \\
34980 & Nashville-Davidson--Murfreesboro--Franklin, TN & 1249 & 2975 & 0.098068 & 0.705842 & 0.041068 & 0.453458 \\
47260 & Virginia Beach-Norfolk-Newport News, VA-NC & 1959 & 5213 & 0.088525 & 0.737602 & 0.055929 & 0.334743 \\
39300 & Providence-Warwick, RI-MA & 1680 & 4462 & 0.105542 & 0.751966 & 0.037493 & 0.559642 \\
33340 & Milwaukee-Waukesha-West Allis, WI & 1922 & 4942 & 0.087781 & 0.615313 & 0.021265 & 0.245400 \\
27260 & Jacksonville, FL & 1417 & 3581 & 0.088426 & 0.708114 & 0.036279 & 0.461129 \\
36420 & Oklahoma City, OK & 2625 & 7366 & 0.107396 & 0.616557 & 0.017379 & 0.224385 \\
32820 & Memphis, TN-MS-AR & 1723 & 4301 & 0.081761 & 0.716835 & 0.039208 & 0.443304 \\
39580 & Raleigh, NC & 2061 & 5413 & 0.105767 & 0.645926 & 0.023171 & 0.298301 \\
40060 & Richmond, VA & 2137 & 5564 & 0.086185 & 0.659416 & 0.030028 & 0.297409 \\
31140 & Louisville/Jefferson County, KY-IN & 1770 & 4664 & 0.085449 & 0.703791 & 0.034157 & 0.392355 \\
35380 & New Orleans-Metairie, LA & 2909 & 7514 & 0.088600 & 0.601389 & 0.020339 & 0.162278 \\
25540 & Hartford-West Hartford-East Hartford, CT & 1332 & 3342 & 0.103619 & 0.631258 & 0.045534 & 0.331053 \\
41620 & Salt Lake City, UT & 2595 & 6227 & 0.091808 & 0.776648 & 0.026630 & 0.465922 \\
13820 & Birmingham-Hoover, AL & 2275 & 6197 & 0.097283 & 0.620737 & 0.022107 & 0.282488 \\
15380 & Buffalo-Cheektowaga-Niagara Falls, NY & 2155 & 5922 & 0.089330 & 0.635451 & 0.023827 & 0.387595 \\
40380 & Rochester, NY & 1468 & 3924 & 0.102368 & 0.696740 & 0.033079 & 0.436049 \\
24340 & Grand Rapids-Wyoming, MI & 2342 & 6639 & 0.095191 & 0.601011 & 0.018051 & 0.266500 \\
\bottomrule
\end{tabular}}

\caption[Topological features of the $52$ metropolitan statistical areas considered in our study.]{\textcolor{black}{\textbf{Topological features of the $52$ metropolitan statistical areas considered in our study.} 
The road network sample $G$ of each metropolitan area has $N$ nodes (intersections) and $K$ edges (roads connecting intersections).
For road network $G$, the inequalities in the distribution of these centralities are captured by the Gini coefficients of closeness ($g^C$), betweeness ($g^B$), straightness ($g^S$), and information ($g^I$).
}}
\label{table_topolical_features}
\end{table}

\end{document}